\documentclass[final,5p,times,twocolumn,superscriptaddress]{revtex4}

\usepackage[ansinew,latin1]{inputenc}
\usepackage{psfrag}
\usepackage{amssymb}
\usepackage{amsmath}
\usepackage{graphicx}

\begin{document}

\title{Testing isotropy of the universe using the Ramsey resonance technique on ultracold neutron spins}

\def\TUM{Technische Universit\"at M\"unchen, D--85748 Garching, Germany}
\def\LPC{LPC Caen, ENSICAEN, Universit\'e de Caen, CNRS/IN2P3, F--14050 Caen, France}
\def\JENA{Department of Neurology, Friedrich--Schiller--University, Jena, Germany}
\def\JUC{Marian Smoluchowski Institute of Physics, Jagiellonian University, 30--059 Cracow, Poland}
\def\ECTUM{Excellence Cluster `Universe', Technische Universit\"at M\"unchen, D--85748 Garching, Germany}
\def\PSI{Paul Scherrer Institut (PSI), CH--5232 Villigen PSI, Switzerland}
\def\PGUM{Institut f\"ur Physik, Johannes--Gutenberg--Universit\"at, D--55128 Mainz, Germany}
\def\JINR{JINR, 141980 Dubna, Moscow region, Russia}
\def\ETH{ETH Z\"urich, CH-8093 Z\"urich, Switzerland}
\def\UNIFR{University of Fribourg, CH--1700, Fribourg, Switzerland}
\def\HNINP{Henryk Niedwodnicza\'nski Institute for Nuclear Physics, 31--342 Cracow, Poland}
\def\LPSC{LPSC, Universit\'e Joseph Fourier Grenoble 1, CNRS/IN2P3, Institut National Polytechnique de Grenoble 53, F--38026 Grenoble Cedex, France}
\def\KCGUM{Institut f\"ur Kernchemie, Johannes--Gutenberg--Universit\"at, D--55128 Mainz, Germany}
\def\KULEUVEN{Instituut voor Kern-- en Stralingsfysica, Katholieke~Universiteit~Leuven, B--3001 Leuven, Belgium}

\author{I.~Altarev}            \affiliation{\TUM}
\author{G.~Ban}                \affiliation{\LPC}
\author{G.~Bison}              \affiliation{\JENA}
\author{K.~Bodek}              \affiliation{\JUC}
\author{M.~Daum}               \affiliation{\ECTUM}\affiliation{\PSI}
\author{M.~Fertl}              \affiliation{\PSI}
\author{P.~Fierlinger}         \affiliation{\ECTUM}
\author{B.~Franke}             \affiliation{\ECTUM}\affiliation{\PSI}
\author{E.~Gutsmiedl}          \affiliation{\TUM}
\author{W.~Heil}               \affiliation{\PGUM}
\author{R.~Henneck}            \affiliation{\PSI}
\author{M.~Horras}             \affiliation{\ECTUM}\affiliation{\PSI}
\author{N.~Khomutov}           \affiliation{\JINR}
\author{K.~Kirch}              \affiliation{\PSI}\affiliation{\ETH}
\author{S.~Kistryn}            \affiliation{\JUC}
\author{A.~Kraft}              \affiliation{\PGUM}
\author{A.~Knecht}             \affiliation{\PSI}
\author{P.~Knowles}            \affiliation{\UNIFR}
\author{A.~Kozela}             \affiliation{\HNINP}
\author{T.~Lauer}              \affiliation{\KCGUM}
\author{B.~Lauss}              \affiliation{\PSI}
\author{T.~Lefort}             \affiliation{\LPC}
\author{Y.~Lemi\`ere}          \affiliation{\LPC}
\author{A.~Mtchedlishvili}     \affiliation{\PSI}
\author{O.~Naviliat-Cuncic}    \affiliation{\LPC}
\author{A.~Pazgalev}           \affiliation{\UNIFR}
\author{G.~Petzoldt}           \affiliation{\ECTUM}
\author{F.~M.~Piegsa}          \affiliation{\ETH}
\author{E.~Pierre}             \affiliation{\LPC}\affiliation{\PSI}
\author{\underline{G.~Pignol}} \affiliation{\ECTUM}
\author{G.~Qu\'em\'ener}       \affiliation{\LPC}
\author{M.~Rebetez}            \affiliation{\UNIFR}
\author{D.~Rebreyend}          \affiliation{\LPSC}
\author{S.~Roccia}             \affiliation{\KULEUVEN}
\author{P.~Schmidt-Wellenburg} \affiliation{\PSI}
\author{N.~Severijns}          \affiliation{\KULEUVEN}
\author{Yu.~Sobolev}           \affiliation{\PGUM}
\author{A.~Weis}               \affiliation{\UNIFR}
\author{J.~Zejma}              \affiliation{\JUC}
\author{J.~Zenner}             \affiliation{\KCGUM}
\author{G.~Zsigmond}           \affiliation{\PSI}

\begin{abstract}
Physics at the Planck scale could be revealed by looking for tiny violations of fundamental symmetries in low energy experiments. 
In 2008, a sensitive test of the isotropy of the Universe using has been performed with stored ultracold neutrons (UCN), 
this is the first clock-comparison experiment performed with free neutrons. 
During several days we monitored the Larmor frequency of neutron spins in a weak magnetic field using the Ramsey resonance technique. 
An non-zero cosmic axial field, violating rotational symmetry, would induce a daily variation of the precession frequency. Our null result constitutes one of the most stringent tests of Lorentz invariance to date.
\end{abstract}


\maketitle

\section{Introduction}
\label{Introduction}

The concept of symmetry plays a central role in our current understanding of physics at its most fundamental level. 
The invariance of physical laws under rotation symmetry is perhaps the most basic example, and for such it is worthwhile to test its validity with high accuracy. 
Moreover, it has been argued that a breakdown of rotation symmetry, or more generally Lorentz symmetry, could arise from quantum gravity effects. 
The high level of precision reached by modern tests of rotation symmetry opens the possibility to observe tiny manifestations of the physics at the Planck scale. 
Here we describe an experiment performed recently with ultracold neutrons based on spin dynamics \cite{Altarev:2009wd}. 
Picture a neutron sitting in an environment free of magnetic field (and electric field). 
One consequence of rotation invariance is the fact that the neutron spin should be stationary, since there is no direction around which it could turn around. 
If the neutron spin were rotating around a direction ${\bf b}$ at a nonzero frequency $\frac{1}{\pi \hbar} b$ ($\hbar$ is the Planck constant), we would observe a breakdown of the rotation invariance. 
Such an effect is parametrized by the exotic potential $V = {\bf \sigma} \cdot {\bf b}$, where ${\bf \sigma}$ are the Pauli matrices acting on the neutron spin states. 
The vector ${\bf b}$, so-called \emph{cosmic spin anisotropy field}, defines a priviliged direction in the universe. 
Now if a neutron is subjected to both a static magnetic field ${\bf B}$ and the new field ${\bf b}$, its spin will precess at the modified Larmor frequency $f_n$, which to first order in $b$ is given by:
\begin{equation}
f_n = \frac{\gamma_n}{2 \pi} B + \frac{1}{\pi \hbar} {\bf b} \cdot \frac{\bf B}{B}.
\end{equation}
If we keep the magnetic field ${\bf B}$ static in the laboratory frame fixed on Earth, say vertical, then the scalar product ${\bf b} \cdot {\bf B}$ will be daily modulated as the Earth is rotating. 
Searching for a daily modulation in the Larmor frequency in a static magnetic field probes the component $b_\bot$ of ${\bf b}$ orthogonal to the Earth's rotation axis. 

The presented result is in fact a \emph{clock comparison experiment} in that the neutron Larmor frequency is compared to the precession frequency of a co-habiting $^{199}$Hg magnetometer. 
In the last section of this article we will discuss our result in regard with the numerous previous clock comparison experiments, in particular those performed with $^{199}$Hg. 
The first clock comparison experiments were performed by Hughes \cite{Hughes} and Drever \cite{Drever}. 
They were motivated by testing Mach's principle, aiming at detecting a dependence of the Zeeman energy of nuclei as a function of the relative orientation of the direction of the magnetic field and the direction to the galactic center. 
Nowadays these searches follow the development of a general formalism for Lorentz violation called the Standard Model Extension (SME) \cite{Kostelecky} that parametrizes all Lorentz-violating effects possibly induced by the yet unknown Planck scale physics. 
The SME contains a number of terms that violate local Lorentz invariance, among which is the cosmic axial field ${\bf b}$. 

Experimental setups to resolve changes in neutron Larmor precession frequency have been in constant development over the past 50 years with the goal of detecting a non zero neutron electric dipole moment. 
For this test of rotation symmetry we used the most sentive one of such apparatus available today, namely the OILL apparatus.

\begin{figure}[h]
\includegraphics[width=1.0\linewidth]{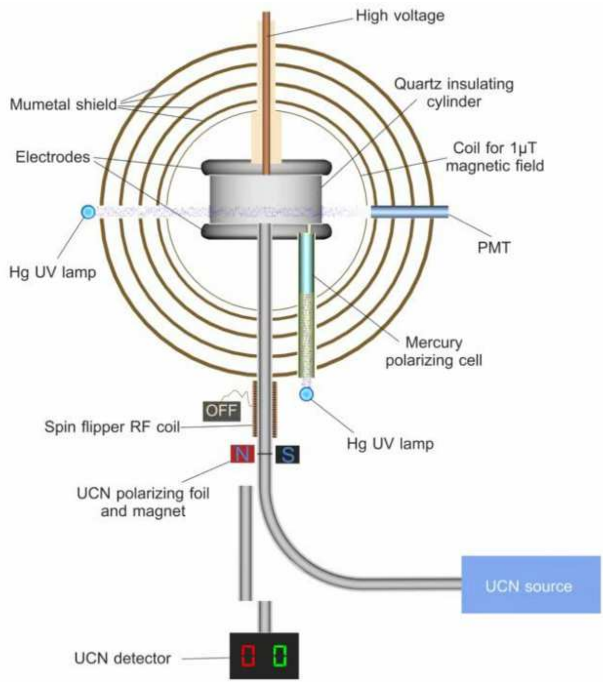}
\caption{Sketch of the OILL apparatus.
} \label{OILLPict}
\end{figure}

\section{The OILL spectrometer}
\label{OILL}

The OILL spectrometer has been built and operated at the PF2 beamline at ILL by the Sussex/RAL/ILL collaboration and it holds the world record limit on the neutron electric dipole moment \cite{Baker}. 
It is now operated by an European collaboration in view of a more sensitive measurement \cite{Altarev2009} at the Paul Scherrer Institut where it will profit from a more intense ultracold neutron source. 
The data reported in this article has been collected at the ILL in 2008, the year preceding the move of the apparatus. 
Figure \ref{OILLPict} shows schematically the OILL spectrometer. 

\subsection{Storing polarized ultracold neutrons}

\begin{figure}[t]
\includegraphics[width=1.0\linewidth]{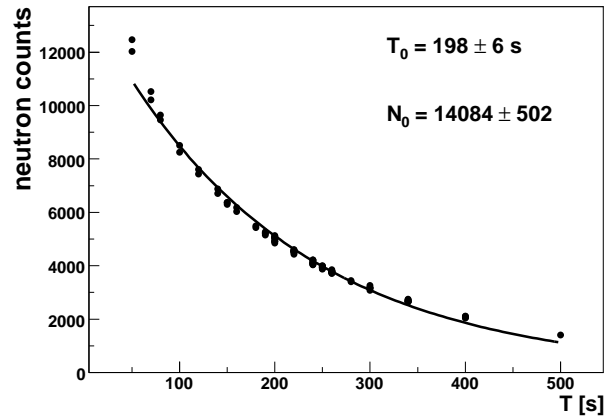}
\caption{Number of UCN counts in the detector as a function of the storage time in the OILL chamber. 
} \label{storage}
\end{figure}

The OILL apparatus uses ultracold neutrons, neutrons with kinetic energy of about $100$~neV. 
Those neutrons are reflected at all angles of incidence by most materials and can be stored in material bottles for a very long time. 
In a typical cycle the UCNs are guided from the source to a storage volume (a cylinder, 12~cm height and 47~cm diameter), filling the volume for about 40~s before closing the UCN valve. 
On the way to the storage volume they are polarized when passing through a magnetized ferromagnetic foil. 
After the completion of the Ramsey procedure described below, the UCN valve is opened again and the neutrons fall down into a He detector where they are counted. 
On the way to the detector they pass again through the magnetized foil that serves as a spin analyser. 
Figure \ref{storage} shows the number of UCN counts as a function of the duration of the storage time. 
From the nearly exponential decrease of the neutron count rate we extract the UCN storage time of $T_0 \approx 200$~s in the OILL chamber. 
The storage curve departs from a simple exponential behaviour because different parts of the UCN energy spectrum have different storage times. 

\subsection{Measuring the neutron Larmor frequency}

\begin{figure}
\includegraphics[width=1.0\linewidth]{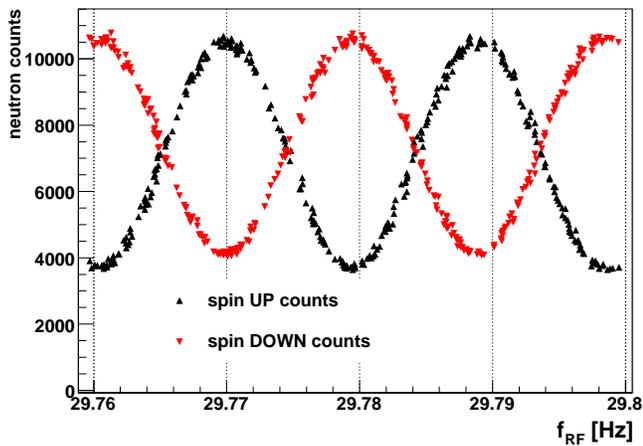}
\caption{Ramsey resonance measured with OILL. Red: counts of the spin up neutrons, black: counts of the spin down neutrons. 
} \label{Ramsey}
\end{figure}

The chamber is exposed to a static vertical magnetic field of $B_0 = 1 \ \mu$T, corresponding to a Larmor frequency of $f_n \approx 30$~Hz. 
During the storage of polarized UCN the Ramsey method of separated oscillatory fields is applied, in order to measure $f_n$ accurately. 
An initial oscillating horizontal field pulse of frequency $f_{\rm RF}$ is applied for 2~s. 
It flips the neutron spin by $\pi/2$. 
Then the UCN spins precess freely around the $B_0$ field, for typically $T = 100$~s. 
A second $\pi/2$ pulse, in phase with the first pulse, is then applied. 
The Ramsey procedure is resonant. 
It flips the neutron spin by $\pi$ only when $f_{\rm RF} = f_n$. 
Figure \ref{Ramsey} shows a measured Ramsey resonance: the vertical component of UCN spins as a function  of the pulse frequency $f_{\rm RF}$. 
To optimize the sensitivity to a change in the Larmor frequency we operate at $f_{\rm RF}$ corresponding to equal spin up and spin down counts in the sides of the central Ramsey fringe where the slope is the steepest. 
One can show that each cycle gives a measurement of the Larmor frequency $f_n$ with a statistical accuracy of: 
\begin{equation}
\label{statAcc}
\sigma_{f_n} = \frac{1}{2 \pi \alpha \ T \sqrt{N} }, 
\end{equation}
where $N$ is the number of neutron counts, and $\alpha$ is the polarization of the neutrons at the end of the storage period. 
In operating conditions this was of the order of $\sigma_{f_n} \approx 30 \ \mu$Hz. 

\subsection{Controlling the magnetic field}
\label{Magnetometry}

\begin{figure}
\includegraphics[width=1.0\linewidth]{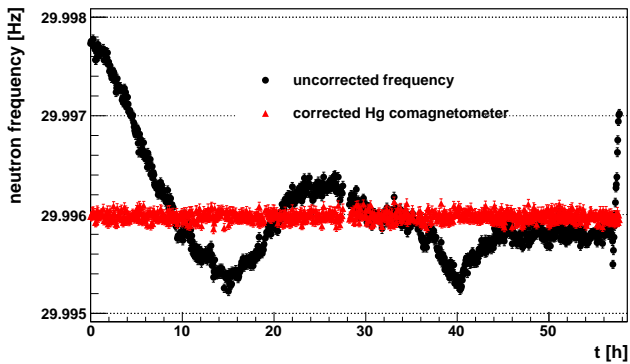}
\caption{Each point is a measurement cycle of the neutron Larmor frequency, uncorrected (dots) and corrected for the B field fluctuations with the comagnetometer (triangles).   
} \label{comagnetometer}
\end{figure}

To resolve a daily modulation in the neutron Larmor frequency, one has to make sure that the magnetic field $B_0$ is not fluctuating too much. 
The control of the magnetic field is the most critical issue for nEDM spectrometers and involves a great deal of shielding, shaping and monitoring of the magnetic field. 
First, the storage chamber is surrounded by a four layer mumetal magnetic shield that protects the experiment against variations of the external field. 
The shielding factor against slow changes of the outside vertical field was measured to be about $10000$. 
But this is far from being enough to lower the inner field fluctuations below the statistical accuracy (\ref{statAcc}) of the neutron cycles, as is illustrated in fig. \ref{comagnetometer}. 
To gain control over the residual magnetic field fluctuations, a unique feature of the OILL spectrometer is the mercury comagnetometer \cite{Green}. 
Within the neutron storage chamber, nuclear spin-polarized $^{199}$Hg atoms precess in the same magnetic field as the neutrons. 
The Larmor frequency $f_{\rm Hg} \approx 8$~Hz of the mercury atoms is measured optically, by recording the oscillation of the transmission of polarized resonant light. 
The mercury comagnetometer provides a field average measurement for each cycle, with the same time average as for the neutrons, at an accuracy of $\sigma_{f_{\rm Hg}} \approx 1 \ \mu$Hz. 
The performance of such a method is illustrated in fig. \ref{comagnetometer} where the neutron frequency is corrected for changes of the magnetic field using the comagnetometer. 
The fluctuations of the corrected neutron frequency are in
agreement with the purely statistical precision (\ref{statAcc}). 

\section{Results and statistical analysis}
\label{Results}

\begin{figure}
\includegraphics[width=1.0\linewidth]{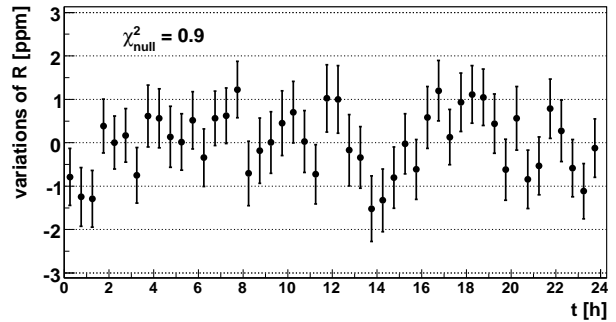}
\caption{Variations of the $R$ ratio around its average. For clarity the data are folded modulo 23.9345 hours and binned every half hour.
} \label{binned}
\end{figure}

The fact that the fluctuations of the magnetic field are suppressed below the statistical sensitivity was used to search for a daily modulation of the neutron Larmor frequency. 
For each measurement cycle we form the ratio of the neutron to mercury Larmor frequency $R = f_n/f_{\rm Hg}$. 
This ratio is free from magnetic field fluctuations. 
Then we parametrize the signal for the daily modulation by the amplitude $A$ and the phase $\phi$: 
\begin{equation}
\label{Rt}
R(t) = \frac{f_n(t)}{f_{\rm Hg}(t)} = \left| \frac{\gamma_n}{\gamma_{\rm Hg}} \right| + A \sin(\Omega t + \phi) + \delta R,
\end{equation}
where $\Omega = 2 \pi /23.9345 \, {\rm h}$ is the sidereal angular frequency, $\gamma_{n}, \gamma_{\rm Hg}$ are the gyromagnetic ratio of the neutron and the mercury atoms, and $\delta R$ is a constant systematic shift that can occur for example due to magnetic field gradients, discussed in the next section. 

In 2008 two series of runs have been conducted. 
The first series, recorded in April-May 2008 contains in total 4.6 days of data in three different continuous runs. 
This series was analysed \cite{Altarev:2009wd} and no daily modulation was found in the ratio $R$. 
An upper limit on the maximum amplitude of the modulation was reported $A < 0.58 \times 10^{-6}$~(95 \% C.L.) using a frequentist statistical method. 
In magnetic field units this would correspond to $B_A = A B_0 \gamma_{\rm Hg}/\gamma_n = 150$~fT. 
A second series was recorded in December 2008, consisting of three continuous runs of 5.6 days in total. 
In this series an electric field was applied, with reversed direction every 2~hours for a nEDM measurement. 
Since the frequency of the E field reversal is large compared to the sidereal frequency the electric field does not affect this analysis even if the neutron EDM were enormous. 

In this article we present a combined analysis of both series, searching for a daily modulation. 
For each run the average value of $R$ has been subtracted, resulting in a subtracted value $\Delta R$ for each cycle, with a standard error $\sigma_{R}$. 
Fig. \ref{binned} shows an overview of all data, folded modulo $23.9345 \, {\rm h}$. The error bars indicate combined statistical errors of the neutron and the Hg frequency, dominated by the former one. 
The whole dataset is compatible with a signal of null amplitude (reduced $\chi^2_{\rm null} = 0.90$). 

\begin{figure}
\includegraphics[width=1.0\linewidth]{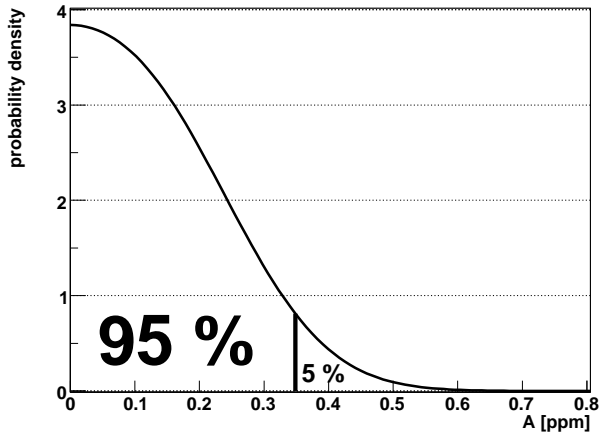}
\caption{ Bayesian probability density for the amplitude $A$. 
} \label{bayes}
\end{figure}

A Bayesian analysis was applied to the data to search for a time variation $R(t)$, Eq. (\ref{Rt}). 
First, the following Chi squared function is established: 
\begin{equation}
\chi^2(A, \phi) = \sum_{i=1}^{3563} \left( \frac{\Delta R_i - A \sin(\Omega t_i + \phi)}{\sigma_{R_i}} \right)^2, 
\end{equation}
where the sum runs over the data cycles. 
The posterior probability density for $A$ is given by the likelihood function: 
\begin{equation}
L(A) = \frac{1}{N} \int_0^{2 \pi} \exp(-\chi^2(A, \phi)) \, d\phi, 
\end{equation}
where $N$ is a normalisation coefficient. 
This function is plotted in Fig. \ref{bayes} from which we extract the following bound:
\begin{equation}
A < 0.35 \times 10^{-6} \quad 95 \ \% \ {\rm C. L.} \quad {\rm (stat)}
\end{equation}
It improves the bound obtained with the first series alone as expected.

\section{Possible systematic effect}
\label{Systematics}

\begin{figure}
\includegraphics[width=1.0\linewidth]{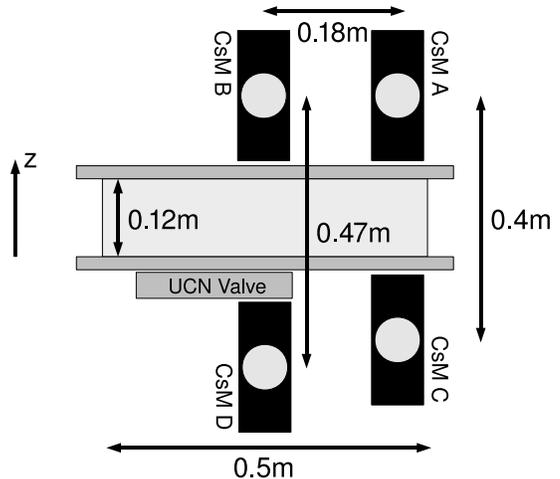}
\caption{
Vertical cut through the cylindrical storage chamber for UCN and $^{199}$Hg. Schematically indicated are the $\approx 70$~mm diameter Cs vapor filled bulbs and their mounts. 
The scalar Cs magnetometer measures the magnitude of $B$ found at the center of the spherical bulb.   
} \label{cesium}
\end{figure}

A systematic effect could arise due to imperfect monitoring of the magnetic field by the comagnetometer. 
If the source of this imperfection is daily modulated it could in principle hide a true cosmic signal. 
The only known relevant effect is the so-called gravitational shift. 
The mercury atoms form a gas at room temperature and fill the storage volume with a uniform density. 
On the contrary, the UCN gas is affected by gravity and the UCN density is significantly higher at the bottom of the storage chamber than at the top. 
This results in a difference in center of mass height $h$ of the two species, that in turn results in a shift of the ratio $R$ in the presence of a vertical gradient $\partial B/ \partial z$ of the magnetic field: 
\begin{equation}
\label{gravShift}
R = \left|\frac{\gamma_n}{\gamma_{\rm Hg}} \right| \left(1 + \frac{h}{B} \, \frac{\partial B}{\partial z} \right)
\end{equation}

A first estimate of $h$ can be obtained from the kinetic theory of gases. 
The maximum energy of stored UCN in the chamber is the Fermi potential of the weakest wall, i.e. $V_F = 95$~neV for the quartz insulator ring. 
We can assign a temperature $T_{\rm UCN}$ to the UCN gas by the matching of the mean energy: $\frac{3}{2} k_B T_{\rm UCN} = \frac{2}{3} V_F$, that is, $T_{\rm UCN} = 0.5$~mK. 
Assuming "thermal" equilibrium for the UCN gas the distribution of height can be derived together with the center of mass offset:
: 
\begin{eqnarray}
f(z) dz &  = &  \exp \left( -\frac{mgz}{k_BT_{\rm UCN}} \right) dz, \\
\label{gasTheory}
h & = & \frac{\int_0^H (H/2-z) f(z) dz}{\int_0^H f(z) dz} = 2.9 \, {\rm mm}, 
\end{eqnarray}
where $H = 12$~cm is the height of the storage chamber. 
This is only an estimate since the UCN energies are certainly not distributed according to the Maxwell-Boltzmann distribution. 

\begin{figure}
\includegraphics[width=1.0\linewidth]{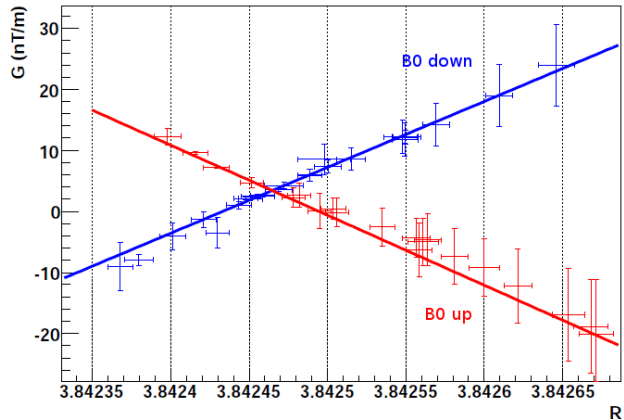}
\caption{Correlation between the vertical field gradient measured by the Cesium magnetometers and the $R = f_n/f_{\rm Hg}$ ratio, for magnetic field pointing downwards (blue) and upwards (red). 
} \label{Rcurve}
\end{figure}

A dedicated experiment was performed to directly observe the gravitational shift (\ref{gravShift}) using four Cesium magnetometers \cite{Groeger} running in parallel with the neutron and mercury systems. 
These magnetometers were placed as described in fig. \ref{cesium}, two above the storage chamber and two below, allowing to measure the magnetic field gradient $\partial B / \partial z$. 
The magnetic field gradient could be adjusted using correction coils at the top and at the bottom of the vacuum chamber. 
Figure \ref{Rcurve} shows the correlation between the gradient extracted from the Cs magnetometers and the ratio $R$. 
Each point corresponds to a different magnetic field configuration, i.e. different currents in the correction coils. 
The correlation is that expected from (\ref{gravShift}), with a measured center of gravity offset of $h = 2.3 \pm 0.1$~mm. 
This is in fairly good agreement with the naive estimate (\ref{gasTheory}). 

\begin{figure}
\includegraphics[width=1.0\linewidth]{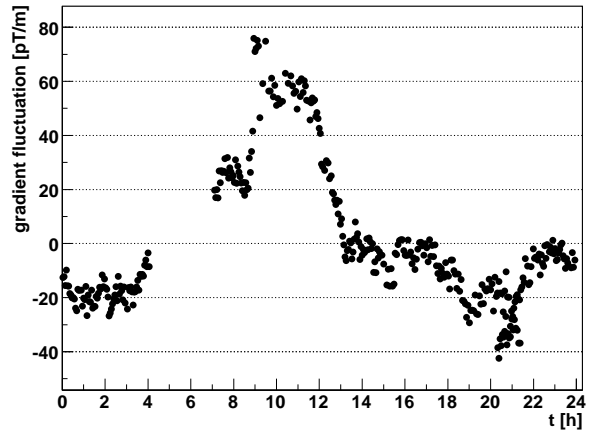}
\caption{Fluctuations of the magnetic field gradient measured by the Cesium magnetometers. 
} \label{gradients}
\end{figure}

A daily modulation of the magnetic field gradient would induce a false modulation of the $R$ ratio through eq. (\ref{gravShift}). 
During part of the datataking, the gradients were measured online with the Cesium magnetometers, as shown in fig. \ref{gradients}. 
From this measurement we extracted the daily modulated Fourier component of this signal, with an amplitude of $30$~pT/m. 
The statistical analysis presented in the previous section was repeated, this time with a modulation of amplitude $0.3$~ppm superimposed to the data. 
The worse case scenario (the worse phase of the superimposed oscillation) leads to the corrected bound: 
\begin{equation}
\label{Astatsys}
A < 0.4 \times 10^{-6} \quad 95 \ \% \ {\rm C. L.} \quad {\rm (stat+syst)}. 
\end{equation}
The systematic effect just discussed is still negligible, but would become an issue for future experiments at the Paul Scherrer Institut with improved statistics. 

\section{Discussion}
\label{discussion}

The result (\ref{Astatsys}) can be interpreted in terms of a limit on the cosmic spin anisotropy field ${\bf b}$ for the free neutron. 
The amplitude $A$ is related to the component $b_\bot$ of ${\bf b}$ orthogonal to the Earth rotation axis: 
\begin{equation}
A = b_\bot \frac{\cos(\lambda)}{\pi \hbar f_{\rm Hg}}
\end{equation}
where $\lambda$ is the latitude of the experiment. 
Thus we extract a limit on the small energy scale associated to the breakdown of rotation symmetry: 
\begin{equation}
\label{result}
b_\bot < 1 \times 10^{-20} \, {\rm eV} \quad 95 \% \, {\rm C.L.}
\end{equation}
This limit constitutes a improvement by a factor of two compared to our previous analysis \cite{Altarev:2009wd} based on only part of the data ($b_\bot < 2 \times 10^{-20}$~eV). 
The new dataset added (December 2008) to the analysis was taken 7 months apart from the first dataset (April-May 2008). 
This is ideal to cancel any day-night fluctuations because the day-night phase and the sidereal phase are in $\pi$ phase opposition after 6 months. 
In this respect, the new combined analysis (\ref{result}) is free of unknown day-night influence. 

Table \ref{ALLResults} compares the present result (\ref{result}) to existing limits on other particles. 
The result reported here is the best limit for the free neutron. 
It is complementary to the more precise atomic experiments \cite{Lamoreaux1995,Bear2000} that can be interpreted as limits concerning bound neutrons inside nuclei. 
Our result using UCN is free from model dependent nuclear corrections 
and related possible suppression effects.
In our analysis we have assumed that the Larmor frequency of the mercury atoms was not daily fluctuating due to the cosmic spin anisotropy field. 
This assumption is justified by the existing limit \cite{Lamoreaux1995} for $^{199}$Hg. 

\begin{table}[hhh]
\begin{center}
\begin{tabular}{lccc}
\hline
Reference	 				&System 	& Particle 		 & $b_{\bot}$ [eV]     \\
\hline 
Berglund \textit{et al.}, \cite{Lamoreaux1995}	& Hg $\&$ Cs 	& bound neutron          & $9 \times 10^{-22}$ \\
						& 	 	& electron               & $2 \times 10^{-20}$ \\
Bear \textit{et al.}, \cite{Bear2000}		& Xe $\&$ He	& bound neutron          & $2 \times 10^{-22}$ \\
Phillips \textit{et al.},\cite{Phillips}	& H		& proton		 & $4 \times 10^{-18}$ \\
Heckel \textit{et al.},	\cite{Heckel:2006ww}    & e		& electron               & $7 \times 10^{-22}$ \\
Bennet \textit{et al.},  \cite{muon}         	& $\mu$         & positive muon          & $2 \times 10^{-15}$ \\
                                                &               & negative muon          & $3 \times 10^{-15}$ \\
{\bf This analysis}				& n $\&$ Hg	& free neutron	 	 & $1 \times 10^{-20}$ \\ 
\hline
\end{tabular}
\caption{\label{ALLResults} Results of more restricting upper limits (at 95$\%$~C.L.) on 
$b_\bot(e)$, $b_\bot(N)$, $b_\bot(p)$, $b_\bot(\mu)$, $b_\bot(n)$ the couplings between a cosmic spin anisotropy 
field and different particles.}
\end{center}
\end{table}

We have shown that if a neutron would sit in a zero magnetic field environment, the period for the rotation of the spin around a priviledged direction in the universe, if any, would certainly be longer than $2$ days.

\bibliographystyle{model1-num-names}

\begin{thebibliography}{00}

\bibitem{Altarev:2009wd}
  I.~Altarev {\it et al.},
  Phys.\ Rev.\ Lett.\  {\bf 103}, 081602 (2009). 

\bibitem{Hughes}
  V.~W.~Hughes {\it et al.}, 
  Phys. Rev. Lett. {\bf 4}, 342 (1960).

\bibitem{Drever}
  R.~W.~P.~Drever, 
  Philos. Mag. {\bf 6}, 683 (1961).

\bibitem{Kostelecky}
  D.~Colladay and V.~A.~Kostelecky, 
  Phys. Rev. D {\bf 58}, 116002 (1998).

\bibitem{Baker} 
  C.~A.~Baker {\it et al.}, 
  Phys. Rev. Lett. {\bf 97}, 131801 (2006).   


\bibitem{Altarev2009}
I.~Altarev {\it et al.}, 
Nucl. Instrum. Methods Phys. Res., Sect. {\bf A 611}, 133-136 (2009).


\bibitem{Altarev:2010vm}
  I.~Altarev {\it et al.},
  arXiv:1006.4967 [nucl-ex].


\bibitem{Green} 
  K.~Green {\it et al},  
  Nucl. Inst. Meth. in Phys. Res. A  {\bf 404}, 381 (1997). 


\bibitem{Groeger} 
S.~Groeger, A.~S.~Pazgalev, A.~Weis, 
  Appl. Phys {\bf B  80}, 6 (2005).


\bibitem{Lamoreaux1995}

  C.~J.~Berglund {\it et al}, 
  Phys.\ Rev.\ Lett.\  {\bf 75}, 1879 (1995).

  
\bibitem{Bear2000}
  D.~Bear {\it et al}, 
  Phys.\ Rev.\ Lett.\  {\bf 85}, 5038 (2000).  


\bibitem{Phillips}
  D.~F.~Phillips {\it et al.},
  Phys. \ Rev.\ {\bf  D 63}, 111101(R) (2001).
  

\bibitem{Heckel:2006ww}
  B.~R.~Heckel {\it et al.},
  Phys.\ Rev.\ Lett.\  {\bf 97}, 021603 (2006).


\bibitem{muon}
  G.~W.~Bennett {\it et al.}  [Muon (g-2) Collaboration],
  Phys.\ Rev.\ Lett.\  {\bf 100}, 091602 (2008).




\end{thebibliography}

\end{document}